\documentclass[aps,twocolumn,prd,showpacs,nofootinbib]{revtex4}
\usepackage{amsmath}
\usepackage{graphicx}
\usepackage{dcolumn}
\usepackage{bm}
\usepackage{amssymb}
\usepackage{latexsym}

\def\be{\begin{equation}}
\def\ee{\end{equation}}
\def\ba{\begin{eqnarray}}
\def\ea{\end{eqnarray}}

\begin{document}

\title{Primordial
Perturbations Spectra in a Holographic Phase}

\author{Yun-Song Piao}
\affiliation{College of Physical Sciences, Graduate School of
Chinese Academy of Sciences, YuQuan Road 19A, Beijing 100049,
China}

\begin{abstract}

In this paper, we suppose that the universe begins in a
holographic thermal equilibrium phase with the diverged
correlation length, and the phase transition to the radiation
phase of standard cosmology goes with the abrupt reducing of
correlation length. In this case, the primordial perturbations may
be induced by
thermal fluctuations in this holographic phase. 
We calculate the spectra of this holographic primordial
perturbations, and find that the scalar spectrum has a slightly
red tilt and the tensor perturbation amplitude has a moderate
ratio, which may be tested in coming observations. The results
plotted in $r-n_s$ plane is similar to that of large field
inflation models. However, for fixed efolding number, they are
generally in different positions.

\end{abstract}


\maketitle

Recently, it has been shown in Ref. \cite{NBV} that the string
thermodynamic fluctuation may lead to a scale invariant spectrum
of scalar metric perturbation, see Ref. \cite{N, BNPV2, BKSE} for
more details and also Ref. \cite{KKLM} for criticisms. The key
point obtaining the scale invariance lies in the specific heat for
a fixed region be proportional to the area, which is a significant
feature of string thermodynamics \cite{DJNT}.


In Ref. \cite{MSC}, it was noted that the specific heat scaling as
the area may also be induced in a high temperature phase depicted
in terms of holographic principle. The holographic principle
\cite{Hooft, Susskind}, which states that the fundamental degrees
of freedom of a physical system are bound by its surface area, is
generally taken as a fundamental property of the microscopic
theory of quantum gravity. In thermal equilibrium phase with the
temperature ${\cal T}$, which obeying the holographic principle,
to obtain the metric perturbations in various wavelength, one need
to calculate the fluctuations of the energy momentum tensor on
various length $R$, up to the maximal correlation length. For each
value of $R$, the holographic phase means that the space region
bounded by a surface $\sim R^2$ can be described by a finite
number of degrees of freedom given by $\pi R^2/G$. These
fundamental degree of freedom will be expected to have the energy
$\cal T$. In this case the total energy may be written as ${\cal
E}=\pi R^2{\cal T}/G$, which gives the special heat
$c_{R}=\partial {\cal E}/\partial {\cal T}\sim R^2$. Thus
dependent of the existence of an early holographic phase, one
obtain a specific heat scaling as the area, which straightly leads
to the scale invariance of scalar spectrum \cite{MSC}.

This scale invariance suggests that the primordial perturbation
arising from a thermal holographic phase may be able to seed the
structure of observable universe, and as well as lead to the
anisotropies observed in the cosmic microwave background.
Thus it is interesting and also significant to ask what is the
distinct feature of this holographic primordial perturbations,
which is testable in coming observations? This will be done in
this paper. To make the discussions here be model independent as
possible, we will be not constrained to some concrete model,
especially the details of phase transition. We only assume that
the holographic phase has a nearly divergent correlation length,
which is required to assure all interesting modes observed today
are in causal contact before phase transition. The phase
transition to the radiation phase of standard cosmology goes with
the abrupt reducing of correlation length.

The correlation length is generally given by
$l\sim c_sh^{-1} $, where $c_s$ denotes the sound speed of metric
perturbation, and $h$ is the comoving Hubble parameter. Though how
to introduce the rapid change of correlation length of metric
perturbation in a holographic phase is open, it seems that the
divergence of correlation length suggests either $c_s$ diverges or
$h\simeq 0$. The case of $h\simeq 0$ may occur e.g. during bounce,
e.g. see Ref. \cite{V} for a relevant review and also Refs.
\cite{BMS, AKK} for recent examples. 
However, the phase transition relevant to holographic phase dose
not need to involve the bounce of background. Thus we will
consider the case of divergent $c_s$. Though the divergence of
$c_s$ can hardly be understood in Einstein gravity and is
generally expected to have a profound origin, for a
phenomenological discussion we may implemented it in an effective
theory, see Ref. \cite{piaocs1}, in which the perturbation
equation is modified with a varying sound speed, while the
evolution of background equation is not changed and is still that
of Einstein gravity.
Thus here the divergence of the correlation length means that when
${\cal T}\sim {\cal T}_c$, where ${\cal T}_c$ is the critical
temperature of phase transition, $c_s$ rapidly approaches
infinity, while ${\cal T}\ll {\cal T}_c$, it is one.

When $c_s$ is very large, we can have $c_sk\gg h$, which means
that the perturbations are deep in the sound horizon. Thus in the
longitudinal gauge, see Ref. \cite{MFB} for details on the theory
of cosmological perturbations, the (00) equation of metric
perturbation may be reduced to \be c_s^2k^2\Phi \simeq
a^2G\delta\rho, \label{c2k2}\ee where $\Phi$ denotes the scalar
metric perturbation and $G$ is the Newton constant, and the terms
relevant to $h$ has been neglected since $c_sk\gg h$.
Eq.(\ref{c2k2}) is Poisson like, but is in relativistic sense.
From Eq.(\ref{c2k2}), we have the scalar perturbation spectrum \be
{\cal P}_{\Phi}(k)\simeq {a^4G^2\over c_s^4 k^4}{\cal P}_{\delta
\rho}(k)={a^4G^2\over c_s^4 k^4}<\delta \rho^2>
, \label{pk0}\ee where $<\delta \rho^2>$ is the mean square
fluctuation of the energy density, which need to be calculated in
each fixed length scale $R$, up to the physical Hubble scale
$a/h$, since what we introduce is only the nearly divergence of
correlation length of metric perturbation, while that of matter
fluctuation is still limited by Hubble scale. Though here the
physical wavelength $a/k$ of matter fluctuations required to
induce the metric perturbation at present observable scale are sub
sound horizon scale, they are generally super Hubble scale, i.e.
$k<h$, see the red solid lines in Fig.1. Thus in principle we are
not able to deduce the metric perturbation by calculating the
matter fluctuation, since in super Hubble scale the matter
fluctuations has frozen. However, if the sound speed during phase
transition is nearly diverged, the case will be different. In this
case the effective physical wavelength of metric perturbation will
obtain a strong suppress $\sim 1/c_s$ and become $ a/(c_s k)$,
which may be sub Hubble scale well, see the red dashed line in
Fig.1. Thus it seems reasonable to replace $R$ with $a/(c_sk)$ in
$<\delta\rho^2>$
, see Ref. \cite{piaocs1} for a more detailed analysis. In thermal
holographic phase, \ba <\delta\rho^2>(R= {a\over c_sk}) & = &
-{1\over R^6}{\partial \over
\partial \beta}\left({\cal F}+\beta{\partial{\cal F}\over
\partial \beta}\right) \nonumber\\ & = & {\pi{\cal T}^2\over
GR^4}.\label{c1}\ea where $\beta=1/{\cal T}$, and $\cal F$ is the
free energy, which is given as \be {\cal F}={\pi R^2{\cal T}\over
G}\ln{\left({{\cal T}_c\over {\cal T}}\right)}, \label{F}\ee by
the energy $ {\cal E}\equiv {\cal F}+\beta\left({\partial {\cal
F}\over
\partial \beta}\right)$. We substitute Eq.(\ref{c1}) into Eq.(\ref{pk0}),
then can obtain \be {\cal P}_{\Phi}(k) \simeq G{\cal T}^2.
\label{pk1}\ee From Eq.(\ref{pk1}), one can see that the spectrum
of scalar metric perturbation is scale invariant.

\begin{figure}[t]
\begin{center}
\includegraphics[width=7cm]{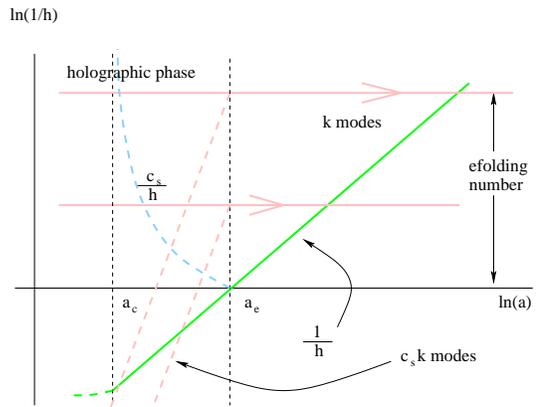}
\caption{The sketch of evolution of the correlation length or
sound horizon $\ln{(c_s/h)}$ and the Hubble horizon $\ln{(1/h)}$
with respect to the scale factor $\ln{a}$. During the transition,
the perturbations are able to leave the sound horizon and become
nearly scale invariant primordial perturbations responsible for
the structure formation of observable universe. }
\end{center}
\end{figure}

When $c_s$ is decreasing, the metric perturbation with some scale
$k^{-1}$ can leave the sound horizon, and quickly freeze. 
The spectrum of curvature perturbation $\xi$ in
comoving supersurface ${\cal P}_{\xi}\simeq {\cal P}_{\Phi}$ up to
a factor with order one, which is constant in the super horizon
scale. Thus the spectrum of the comoving curvature perturbation
can be nearly scale invariant and its amplitude can be calculated
at the time when the perturbation exits the sound horizon, i.e.
$k=h/c_s$, which gives the value of ${\cal T}$ in Eq.(\ref{pk1})
at the sound horizon crossing.

The sound horizon crossing requires $k=h/c_s$, thus we can obtain
\be {k_e\over k} \simeq c_s\equiv\left({{\cal T}_c\over {\cal T}_c
- {\cal T}}\right)^{p},\label{kke1}\ee where the subscript 'e'
denotes the end time of phase transition and thus $c_{s(e)}\simeq
1$, and we also neglected the change of $h$ during phase
transition. When we include the change of $h$, there will be a
factor like $({\cal T}/ {\cal T}_e)^n$ before the right hand term
of Eq.(\ref{kke1}), where $n\sim {\cal O}(1)$, which is negligible
when being compared to that of $c_s$. In Eq.(\ref{kke1}), $c_s$ is
taken by using the analysis in condensed matter physics, where
$p>0$ denotes the critical exponent of transition. By using
Eq.(\ref{kke1}), the spectrum (\ref{pk1}) of metric perturbation
can be rewritten as \be {\cal P}_{\Phi} \simeq G{\cal
T}_c^2\left[1-\left({k\over k_e}\right)^{1/ p}\right]^2. \ee Note
that $k_e$ is the last mode to be generated, thus we have $k<
k_e$, especially on large scale $k\ll k_e$. Thus the amplitude is
approximately ${\cal P}_{\Phi}\simeq G{\cal T}_c^2$. The spectral
index is given by \be n_s -1 = {-2\left({k\over k_e}\right)^{1/
p}\over p\left[ 1-\left({k\over k_e}\right)^{1/ p}\right]}.
\label{ns1}\ee We can see that the scalar spectrum is red and its
tilt is determined by the critical exponent and the temperature at
horizon crossing, since $k$ is related to ${\cal T}$ by
Eq.(\ref{kke1}).

\begin{figure}[t]
\begin{center}
\includegraphics[width=7cm]{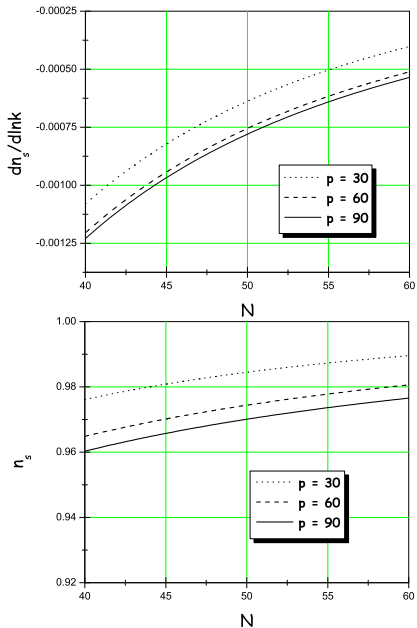}
\caption{The figure of the scalar spectral index $n_s$ and its
running $dn_s/d\ln{k}$ with respect to the efolding number $\cal
N$ for different critical exponent $p$. }
\end{center}
\end{figure}

We define \be {\cal N}\equiv\ln{\left({k_e\over
k}\right)},\label{N}\ee which measures the efolding number of mode
with some scale $\sim k^{-1}$ which leaves the horizon before the
end of phase transition. When taking the comoving Hubble parameter
$h=h_0$, where the subscript `0' denotes the present time, we
generally have ${\cal N}\sim 50$, which is required by observable
cosmology. From Eq.(\ref{kke1}), we can see that when ${\cal
T}\rightarrow {\cal T}_c$, $k_e/k$ nearly approaches to infinity,
thus the efolding number is actually always enough as long as the
initial volume of holographic phase is large enough. We substitute
Eq.(\ref{N}) into Eq.(\ref{ns1}) and can obtain the scalar
spectral index \be n_s-1=-{2\over p}\cdot\left({1\over e^{{\cal
N}/p}-1}\right). \label{ns2}\ee The running of spectral index is
given by \be {dn_s\over d\ln{k}}= {n_s-1\over p}\cdot
\left({1\over 1-e^{-{\cal N}/p}}\right), \label{nrun}\ee which is
the same as that in Ref. \cite{MSC} only when the efolding number
${\cal N}\gg p$. The figure of $n_s$ and $dn_s/d\ln{k}$ with
respect to the efolding number $\cal N$ for different critical
exponent $p$ are plotted in Fig.2. It seems that to fit the
observations \cite{WMAP}, $p$ should be very large. Substituting
Eq.(\ref{ns2}) into Eq.(\ref{nrun}), we can obtain
$dn_s/d\ln{k}\sim 1/p^2$, which means that the running is quite
small, which may also be seen in Fig.2.

Then we will calculate the tensor perturbation spectrum. Though it
is not necessary that the divergence of correlation length of
tensor perturbation is same with that of scalar metric
perturbation, here we assume that they share same one, like
Eq.(\ref{kke1}). In principle the tensor perturbation may be
produced from the vacuum without any source, as happens during
inflation. However, when the source is included, its generation
mechanism will be quite different. In this case the source term
will master the equation of tensor perturbation. When $c_sk\gg h$,
similar to the analysis of scalar metric perturbation, we can find
that in left hand side of the equation of tensor perturbation,
compared with $c_s^2k^2h_{ij}(k)$, other relevant terms with
$h_{ij}$ may be neglected. For example, in unit of Hubble time
$h_{ij}^{\prime\prime}$ may be written as $h^2\Delta h_{ij}$, thus
we have $h_{ij}^{\prime\prime}\simeq h^2\Delta h_{ij} \ll c_s^2k^2
h_{ij}$. Thus the equation of tensor perturbation may be reduced
to \be c_s^2k^2 h_{ij}(k)\simeq a^2G\delta {\cal T}_{ij}(k),
\label{tensor}\ee where $h_{ij}$ is the tensor perturbation and
can be expanded in term of the two basic traceless and symmetric
polarization tensors $e^{\pm}_{ij}$ as $h_{ij} = \sum
h_{\pm}e^{\pm}_{ij}$. Thus we have the tensor perturbation
spectrum \be {\cal P}_{\rm T} \simeq {a^4G^2\over c_s^4
k^4}C_{\,\,j\,\,j}^{i\,\,i}(R= {a\over c_sk}). \ee where
$C_{\,\,j\,\,j}^{i\,\,i}$ is the offdiagonal spatial parts of the
correlation function of energy momentum tensor. However, their
order of magnitude may be estimated by the order of magnitude of
diagonal parts
\be C_{\,\,i\,\,i}^{i\,\,i}={1\over
R^2\beta}{\partial p\over
\partial R}= {\pi{\cal T}^2\over GR^4}\ln\left({{\cal T}_c\over {\cal
T}}\right), \label{c2}\ee which is a good approximation, as was
pointed out in Ref. \cite{BNPV}, where the pressure $p$ can be
obtained by the free energy $p\equiv -{\partial {\cal F}\over
R^3\partial(\ln {R})}$ with Eq.(\ref{F}), see Refs. \cite{BNPV,
BNPV2} for details. Thus we may have \be {\cal P}_{\rm T}\simeq
G{\cal T}^2\ln\left({{\cal T}_c\over {\cal T}}\right).
\label{pt}\ee Note that when ${\cal T}= {\cal T}_c$, $\ln{({\cal
T}_c/{\cal T})}= 0$. However, this does not means that for present
observable scale, ${\cal P}_{\rm T}=0$ since for fixed efolding
number, generally ${\cal T}<{\cal T}_c$, see Fig.1. Thus dependent
on the degree of ${\cal T}\rightarrow {\cal T}_c$, the tensor
perturbation will have a moderately decreasing amplitude, which
means that its spectrum is slightly blue tilt. This can also be
seen as follows \ba {\cal P}_{\rm T}& \simeq & G{\cal
T}_c^2\left[1-\left({k\over k_e}\right)^{1/
p}\right]^2\ln\left[1-\left({k\over k_e}\right)^{1/
p}\right]^{-1}\nonumber\\ &\simeq & G{\cal T}_c^2\left({k\over
k_e}\right)^{1/ p}, \ea where Eq.(\ref{kke1}) and $k\ll k_e$ on
large scale have been used in the first and second lines,
respectively. Thus the spectral index of tensor spectrum $n_{\rm
T}\simeq 1/p$, which is slightly blue. This is similar to the case
in string gas mechanism \cite{BNPV}, but different from that of
slow-roll inflation in which the tilt of tensor spectrum is
generally red.

The ratio $r$ of tensor to scalar is an important quantity for
observation, which as well as $n_s$ makes up of the $r-n_s$ plane
\cite{DKK, K}, in which different classes of inflation modes are
placed in different regions. For the holographic primordial
perturbations, we have the ratio of tensor to scalar
\be r\equiv {{\cal P}_{\rm T}\over {\cal P}_{\xi}}
=C\ln\left({{\cal T}_c\over {\cal T}}\right), \label{r}\ee where
Eqs.(\ref{pk1}) and (\ref{pt}) have been used, and $C\sim {\cal
O}(1)$ is expected to be a constant. Note that Eqs.(\ref{c2k2})
and (\ref{tensor}) are valid only in the approximation $c_sk\gg
h$, thus in principle we need to solve full equation of $\Phi$ and
$h_{ij}$ to determined the value of $C$. Besides, $C$ is also
determined by some exact relations between the correlation lengths
of scalar mode and tensor mode, between the metric perturbation
${\cal P}_{\Phi}$ and the curvature perturbation ${\cal P}_{\xi}$,
between the magnitudes of diagonal spatial parts
$C_{\,\,i\,\,i}^{i\,\,i}$ and offdiagonal spatial parts
$C_{\,\,j\,\,j}^{i\,\,i}$ of the correlation function of energy
momentum tensor. We substitute Eqs.(\ref{kke1}) and (\ref{r}) into
Eq.(\ref{ns1}) and can obtain \be n_s-1 = -{2\over p}\cdot
\left(e^{r/C}-1\right) . \label{nr}\ee Thus for a definite $C$,
the relation between $n_s$ and $r$ is only dependent of the
critical exponent $p$. In principle, for different $C$, one always
may maintain the qualitative behavior of $r-n_s$ lines by changing
the value of $p$. To match it to observations, we can note that
$C>1$ is obviously not favored since this will lead to a quite
large tensor perturbation. Here no loosing generality, we set $C =
0.2$. The $r-n_s$ figure for different critical exponent is
plotted in Fig.3. The black square dots in Fig.3 denote those with
the efolding number ${\cal N}=50$, which can fit recent
observations \cite{WMAP} well, see also Refs. \cite{KKMR} and
recent \cite{CLM} for discussions on the bounds on $r-n_s$.

The lines in Fig.3 lies in the region of large field inflation
models, see e.g. Ref. \cite{LR} for the discussions on various
inflation models and also recent Ref. \cite{AL}, which suggests
that the holographic primordial perturbations may has similar
prediction with large field inflation models. However, they can be
distinguished by the efolding number. For large field inflation
models $\sim \phi^p$, in the slow-roll approximation, for fixed
efolding number, the larger $p$ is, the more the spectral index
$n_s$ deviates from scale invariant. The case with least deviation
corresponds to $p=2$, which is $m^2\phi^2$ model, in which
$n_s-1\simeq -(2/{\cal N})$. While in the holographic primordial
perturbations discussed here, with Eq.(\ref{ns2}), for small $p$,
there will a suppression from $e^{{\cal N}/p}$, thus the maximal
deviation from scale invariant occurs only when ${\cal N}\ll p$.
In this case, we may expand the exponent in the numerator of
Eq.(\ref{ns2}) to the second order ${\cal N}/p$ and obtain \be
n_s-1 \simeq -{2\over {\cal N}}\cdot(1-{{\cal N}\over 2p}).
\ee We can see that the deviation is always small than $-(2/{\cal
N})$. From Eq.(\ref{nr}), ${\cal N}\ll p$ means that to obtain an
enough deviation from scale invariant the ratio $r$ of tensor to
scalar should be very large, which seems be not favored by
observations.

Therefore, generally in the region of large field inflation models
of $r-n_s$ plane, for fixed efolding number, the deviation of
holographic primordial perturbation spectra from scale invariant
is always smaller than that of large field inflation models, which
is independent of the value of $C$. This can be also seen in blue
dashed line in Fig.3, which is plotted by combining
Eqs.(\ref{ns2}) and (\ref{nr}) in which $p$ can cancelled. This
line reflects the relation between $r$ and $n_s$ for fixed
efolding number ${\cal N}$, in which ${\cal N}=50$ has been taken.
This result provides an interesting smoking gun, which means that
for ${\cal N}>50$, see Ref. \cite{LL} for a review on the value of
$\cal N$, if the observations give $n_s<0.96$, then the
holographic primordial perturbations will be ruled out.

\begin{figure}[t]
\begin{center}
\includegraphics[width=7cm]{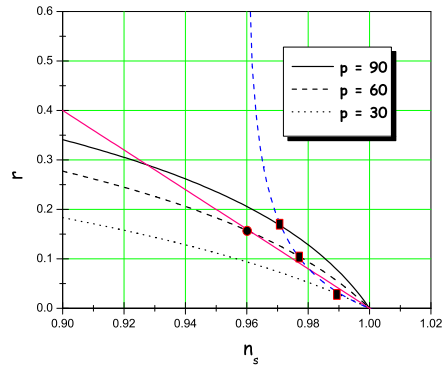}
\caption{The $r-n_s$ figure with $C=0.2$. The blue dashed line
denotes the line with the efolding number ${\cal N}=50$, which
crosses the lines with $p=30, 60, 90$ with the crossing dots
denoted by the black square. The red line correspond to that of
$m^2\phi^2$ inflation model, in which the black round dot denotes
that with ${\cal N}=50$ }
\end{center}
\end{figure}

In summary, we calculate the primordial perturbation spectra
arising from a holographic phase with the assumption that the
correlation length of metric perturbation diverges when the
holographic phase is arrived. The results shown in $r-n_s$ plane
are in the range of large field inflation models with a slightly
red tilt of scalar spectrum and a moderate ratio of tensor to
scalar amplitude. However, for fixed efolding number, they are
generally in different positions. The qualitative features of
perturbation spectra are only dependent of the thermodynamics of
holographic phase and the change of correlation length during
phase transition, but not the details of phase transition. In this
sense the predictions of holographic primordial perturbation given
in this letter are unambiguous, which may be tested in coming
observations.


\textbf{Acknowledgments} This work is supported in part by NNSFC
under Grant No: 10405029, in part by the Scientific Research Fund
of GUCAS(NO.055101BM03), as well as in part by CAS under Grant No:
KJCX3-SYW-N2.

\end{document}